\documentclass[aps,prb,article,twocolumn,showpacs,preprintnumbers,amsmath,amssymb,superscriptaddress]{revtex4}

\date{\today}
\usepackage{epsfig}
\usepackage[toc,page]{appendix}
\usepackage{subfigure}
\usepackage{graphicx}
\usepackage{dcolumn}
\usepackage{bm}
\usepackage{bbm}
\usepackage[colorlinks,linkcolor=blue,hyperindex,CJKbookmarks]{hyperref}
\usepackage{float}
\usepackage{hyperref}
\usepackage{mathtools}
\usepackage{physics}
\usepackage{comment}
\allowdisplaybreaks

\setcitestyle{numbers,square}
\begin{document}

\title{Theory for time-resolved resonant inelastic x-ray scattering}
\author{Yuan Chen}
 \affiliation{Department of Applied Physics, Stanford University, California 94305, USA}
 \affiliation{Stanford Institute for Materials and Energy Sciences, SLAC National Accelerator Laboratory and Stanford University, Menlo Park, California 94025, USA}
\author{Yao Wang }
\affiliation{Department of Applied Physics, Stanford University, California 94305, USA}
 \affiliation{Stanford Institute for Materials and Energy Sciences, SLAC National Accelerator Laboratory and Stanford University, Menlo Park, California 94025, USA}
\affiliation{Department of Physics, Harvard University, Cambridge, MA 02138, USA 
 }
\author{Chunjing Jia}
\affiliation{Stanford Institute for Materials and Energy Sciences, SLAC National Accelerator Laboratory and Stanford University, Menlo Park, California 94025, USA}
\author{Brian Moritz}
\affiliation{Stanford Institute for Materials and Energy Sciences, SLAC National Accelerator Laboratory and Stanford University, Menlo Park, California 94025, USA}
\affiliation{Department of Physics and Astrophysics, University of North Dakota, Grand Forks, North Dakota 58202, USA}
\date{\today}
\author{Andrij M. Shvaika}
\affiliation{Institute for Condensed Matter Physics of the National Academy of Sciences of Ukraine, Lviv 79011, Ukraine}
\author{James K. Freericks}
\affiliation{Department of Physics, Georgetown University, Washington, D.C. 20057, USA}
\author{Thomas P. Devereaux}
 \email[All correspondence should be addressed to T. P. D. (\href{mailto:tpd@stanford.edu}{tpd@stanford.edu})
]{}
\affiliation{Stanford Institute for Materials and Energy Sciences, SLAC National Accelerator Laboratory and Stanford University, Menlo Park, California 94025, USA}
\affiliation{Department of Materials Science and Engineering, Stanford University, California 94305, USA}
\affiliation{Geballe Laboratory for Advanced Materials, Stanford University, California 94305, USA}
\begin{abstract}
Time-resolved measurements of materials provide a wealth of information on quasiparticle dynamics, and have been the focus of optical studies for decades. In this paper, we develop a theory for explicitly evaluating time-resolved resonant inelastic x-ray scattering (tr-RIXS). We apply the theory to a non-interacting electronic system and reveal the particle-hole spectrum and its evolution during the pump pulse. With a high-frequency pump, the frequency and amplitude dependence analysis of the spectra agrees well with the steady state-assumptions and Floquet excitations. When the pump frequency is low, the spectrum extracts real-time dynamics of the particle-hole continuum in momentum space. These results verify the correctness of our theory and demonstrate the breadth of physical problems that tr-RIXS could shed light on. \\
\end{abstract}

\maketitle
\section{Introduction}
Ultrafast materials science, traditionally focused either on time-resolved optical studies or time-resolved angle-resolved photoemission spectroscopy (tr-ARPES), can reveal crucial information about materials' dynamics, including excitation and relaxation of quasiparticles\citep{NuhGedik2010band,Lanzara2012,ZX2012,wang2014real,dal2014snapshots,wang2016using} and phases induced or suppressed by disturbance\citep{ZX2008,Lanzara2011,Light-inducedSuperconduct,lee2012phase,hellmann2012time,kim2012ultrafast,
Cavalleri2016}. The combination of these two techniques provides fine details about the single-particle properties and the long-wavelength collective excitations. However, with an increasing demand for resolving complex excitations, as well as for understanding correlation effects behind them, these two techniques become insufficient in many scenarios. For example, an outstanding issue in condensed matter physics is to detect bosonic excitations across the entire Brillouin zone, as a finite-momentum excitation may be crucial for some ordered phases such as charge density waves, stripes, and superconductivity. Optical techniques such as reflectivity and Raman scattering can only provide information about the resonant excitations at zero momentum\citep{basov2011, TPD2007}, while tr-ARPES detects exclusively single-particle information\citep{ZX2003review}. Even though the collective modes can be inferred qualitatively from tr-ARPES in some cases\citep{ZX2003review, cuk2005review,Lanzara2008, johnston2010systematic,johnston2012evidence}, generally it can only reflect their integrated effects, making it impractical to decipher the full momentum-resolved collective excitations\citep{TPD2016}.

On the other hand, resonant inelastic x-ray scattering (RIXS), as a photon-in photon-out spectroscopy, is increasingly popular due to its capability for detecting a variety of atomically-specified collective excitations in a wide range of momentum and energy\citep{KotaniReview,TPDReview}. Recent development of instrumentation has brought much progress to the energy resolution of RIXS measurements\citep{saxes,Instrumentation2014,Instrumentation2017}. With full Brillouin zone access in momentum space and light polarization selection, RIXS has been able to separate and depict the full momentum-energy structure of  charge, orbital, spin, and lattice degrees of freedom\citep{lu2005charge,ellis2008charge,Ghiringhelli2009MagneticDD,
Dean2012spin,ZX2014dd,moser2015phonon,peng2017}. These advantages make RIXS an indispensable technique for characterizing multiparticle excitations\citep{wang2018theoretical}.


\begin{figure*}[!t]
\begin{center}
\includegraphics[width=16cm]{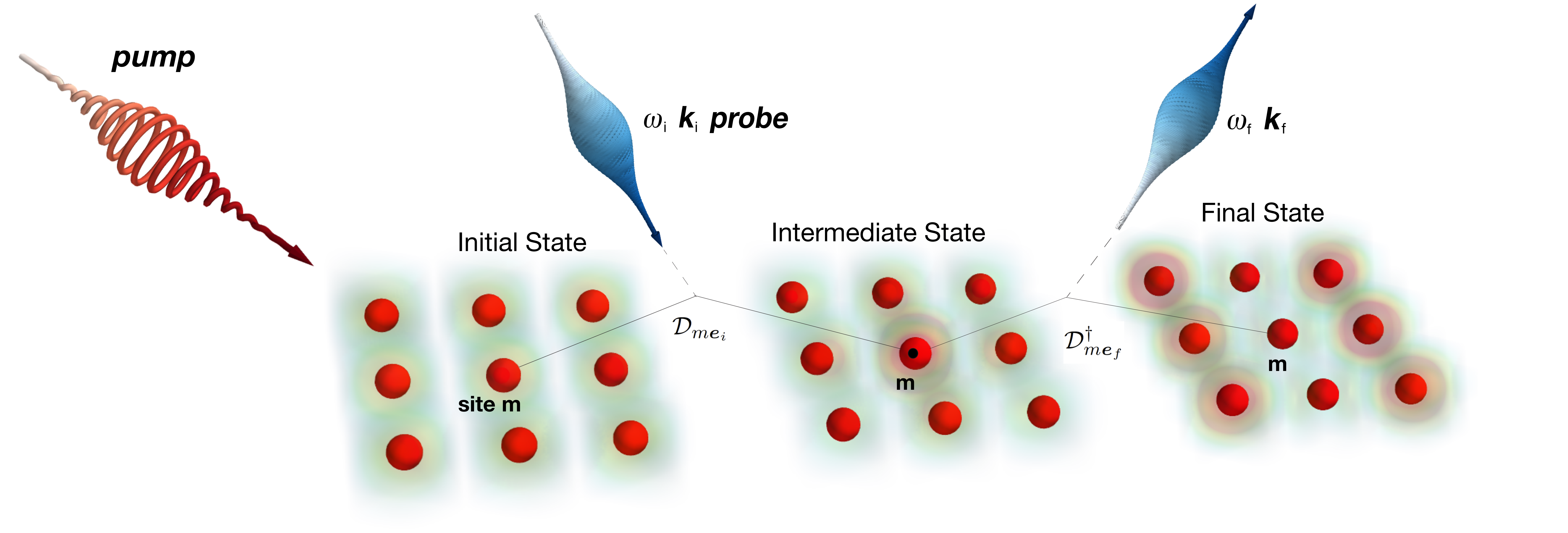}
\caption{\label{fig:0} An illustration of the RIXS process. It has two steps: in the excitation step, the incoming photon with energy $\omega_{\rm i}$ and momentum $\bm{k}_{\rm i}$ promotes an inner shell electron to the outer shell at site $m$ in the initial state and creates the intermediate state. Later, in the de-excitation step, the intermediate state emits the outgoing photon with energy $\omega_{\rm f}$ and momentum $\bm{k}_{\rm f}$ and the core hole is filled, leaving the final state with an excitation.}
\end{center}
\end{figure*}
Recently, progress has been achieved in time-resolved RIXS (tr-RIXS) techniques, with the help of ultrashort and ultrabright x-ray sources\citep{LCLS}. 
Dean et al.\ reported a femtosecond tr-RIXS experiment on $\text{Sr}_2\text{IrO}_4$ to study subtle spin and charge dynamics\citep{Dean2016trRIXS,arxiv_Dean}. Combining the advantages of ultrafast techniques and RIXS, this experiment reveals that the photoinduced suppression of the magnetic order happens mainly for $(\pi, \pi)$ momentum transfers, with in-plane spin correlations restoring on a much faster timescale than the out-of-plane correlations. Mitrano et al. reported tr-RIXS on $\text{La}_{2-x}\text{Ba}_x \text{CuO}_4$ to study evolution of collective modes associated with charge order\citep{Abbamonte2018}. It was found that low energy collective excitations are overdamped and propagate via Brownian-like diffusion, displaying universal scaling behavior arising from the propagation of topological defects. These results have demonstrated a first view of the power of tr-RIXS. With the ability to characterize various multi-particle excitations with time, momentum and energy resolution, tr-RIXS has paved the way for understanding and manipulating nonequilibrium properties.

However, in contrast to the well-established theory in time-resolved single-particle and two-particle studies\citep{FreericksPRL2009,sentef2013prx,review2014nonequilibrium,bonca2014optical,wang2014real} and equilibrium RIXS\citep{KotaniReview,TPDReview}, the theory of tr-RIXS has not yet been developed\citep{wang2018theoretical}. Here, we report a theory for the nonequilibrium tr-RIXS cross-section. While the generic formalism can be applied to arbitrary electronic systems, we explore the theory for non-interacting electrons as a benchmark. By tuning the pump-probe parameters (i.e., pulse width, frequencies, etc.) and tracking the time-dependent spectrum during the pump pulse, we analyze the change of the particle-hole continuum. Floquet replicas and energy renormalizations by photo-induced transient states are clearly detected for high pump frequencies, while breathing and flattening of particle-hole excitations near the momentum nesting point are observed for low pump frequencies. In both cases, tr-RIXS precisely detects the relevant dynamics of particle-hole pairs with full momentum and energy resolution. These results help to set the stage for an understanding of the dynamics for collective excitations in more realistic and strongly correlated systems.

The remaining part of this paper is arranged as follows. In Sec.\ref{sec2} we propose a theory for calculating tr-RIXS as well as its background, derivation, and comparison with the equilibrium formula. In Sec.\ref{num}  we present and analyze numerical results of our theory benchmarked on non-interacting electrons under high- and low-frequency pump pulses. Sec.\ref{diss} closes with a discussion of the relevance of our results for more realistic simulations for correlated systems.

\section{Theory for time-resolved X-ray spectroscopy}
\label{sec2}

RIXS is a second order x-ray scattering process involving resonant intermediate states. In contrast to non-resonant scattering, it has the advantage of strong intensity and large momentum accessibility at the desired electron energy range to which the probe x-ray is tuned\citep{TPDReview}. 
RIXS describes the following photon-in-photon-out process: first, an incoming photon with energy $\omega_{\rm i}$ excites a ground-state core-shell electron to the local valence shell (described by $\mathcal{D}_{m\bm{e}_{\rm i}}$); then, as the second step, an electron from the valence shell annihilates the core hole emitting a photon with energy $\omega_{\rm f}$ (described by $\mathcal{D}_{m\bm{e}_{\rm f}}^{\dagger}$), leaving the system in an excited final state. This is illustrated in Fig.\ref{fig:0}. 
To be self-contained, we first briefly review equilibrium RIXS theory\citep{TPDReview,KotaniReview}. Then we derive the cross-section for tr-RIXS from perturbation theory in the probe pulse. 

\subsection{RIXS Calculation in Equilibrium}
The (equilibrium) RIXS process is a second-order process that consists of two dipole transitions. Its cross section is usually evaluated by the Kramers-Heisenberg formula\citep{KotaniReview,TPDReview}
\begin{equation}\label{KS}
I(\omega_{\rm i},\omega_{\rm f},\bm{q})=\frac{1}{\pi}\Im\bra{\Psi}\frac{1}{\mathcal{H}-E_0-\Delta\omega-i0^+}\ket{\Psi},
\end{equation}
with
\begin{equation}\label{Psi}
\ket{\Psi}=\sum_{m}e^{i\bm{q}\cdot \bm{r}_m}\mathcal{D}_{m\bm{e}_{\rm f}}^\dagger \frac{1}{\mathcal{H}-E_0-\omega_{\rm i}-i\Gamma} \mathcal{D}_{m\bm{e}_{\rm i}}\ket{\Psi_{\rm i}^0}.
\end{equation}
Here $\mathcal{H}$ is the Hamiltonian (including a core-hole interaction), $\omega_{\rm i}, \omega_{\rm f}$ are incident and outgoing photon energies respectively, $\Delta\omega=\omega_{\rm i}-\omega_{\rm f}$ is the energy loss, $\bm{q}$ is the momentum transfer, i.e., the difference between the incoming photon momentum $\bm{q}_{\rm i}$ and the outgoing photon momentum $\bm{q}_{\rm f}$, $\bm{r}_m$ is the position of the $m^{th}$ lattice site, and $E_0$ is the energy of the ground state $\ket{\Psi_{\rm i}^0}$. For direct RIXS, i.e., where the core electron is excited directly into the valence band, $\mathcal{D}_{m\bm{e}}=\sum_{\sigma,\alpha, \beta}M_{\alpha\beta}^{\bm{e}}p_{m\alpha\sigma}^{\dagger}d_{m\beta\sigma}$ is the local dipole transition operator, where $p_{m\alpha\sigma}$ $(d_{m\beta\sigma})$ annihilates an electron at site $m$ with spin $\sigma$ in the valence $\alpha$ (core $\beta$) orbital. We use $M_{\alpha\beta}^{\bm{e}}$ to denote the associated matrix element with photon polarization $\bm{e}$. In the nonequilibrium case, $M_{\alpha\beta}^{\bm{e}}$ will become time-dependent, thus we represent the matrix element and dipole transition operator at time $t$ as $M_{\alpha\beta}^{\bm{e}}(t)$ and $\mathcal{D}_{m\bm{e}}(t)$ respectively. 

\begin{figure}[!t]
\includegraphics[width=8cm]{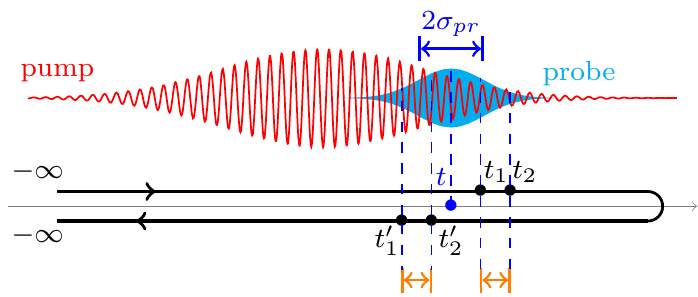}
\caption{\label{fig:1} The Keldysh contour of tr-RIXS. The red curve and cyan shade represent the pump and probe pulse respectively. $t_1$, $t_2$, $t'_2$, and $t'_1$ are as defined in Eq.\eqref{4th-order}. The probe centered at $t$ has a width $2\sigma_{\textrm{pr}}$, containing all four time points. The separations between $t_1(t'_1)$ and $t_2(t'_2)$ correspond to the time when the core hole is present and are further constrained by $l(t_1,t_2)$ or $l(t'_1,t'_2)$, as defined in the main text.}
\end{figure}

\subsection{Time-Resolved RIXS}
To calculate the nonequilibrium time-resolved RIXS cross-section, we assume that the system starts out in equilibrium (i.e., when $t\rightarrow-\infty$) with Hamiltonian $\mathcal{H}_0$. It can then be represented by an ensemble of electron and photon eigenstates $\ket{\Psi_{\rm n}^0}\otimes \ket{\Phi_{\rm ph}}$. The electronic eigenstates $\ket{\Psi_{\rm n}^0}$ satisfy $\mathcal{H}_0\ket{\Psi_{\rm n}^0}=E_{\rm n}\ket{\Psi_{\rm n}^0}$ and are present with probability $\rho_{\rm n}=Z^{-1}\exp[-E_{\rm n}/k_B T]$, where $Z=\sum_{\rm n} \exp[-E_{\rm n}/k_B T]$ is the partition function, and $T$ is the initial temperature. As the linearization of a coherent state at the weak-probe limit, the photon part $\ket{\Phi_{\rm ph}}$ is a one photon state $a^\dagger_{\bm{q}_{\rm i}\bm{e}_{\rm i}}\ket{0}$ of the incoming probe light, where $\ket{0}$ is the photon vacuum and $a^\dagger_{\bm{q}_{\rm i}\bm{e}_{\rm i}}$ is the photon creation operator at incoming probe momentum $\bm{q}_{\rm i}$ with polarization $\bm{e}_{\rm i}$. As the pump is turned on, the Hamiltonian $\mathcal{H}(t)$ becomes time-dependent, and the states evolve according to $\mathcal{U}(t,-\infty)\ket{\Psi_{\rm n}^0}\otimes a^\dagger_{\bm{q}_{\rm i}\bm{e}_{\rm i}}\ket{0}$. Here $\mathcal{U}(t, t')=\mathcal{T}\exp[-\int_{t'}^{t}\mathcal{H}(\tau)d\tau]$ is the time evolution operator and $\mathcal{T}$ is the time-ordering operator. The probe pulse is generally weak compared to the pump and here we treat it as a perturbation $\mathcal{H}_{\textrm{probe}}(t)$, while the pump is exactly included in $\mathcal{H}(t)$, as in the theory for tr-ARPES\citep{FreericksPRL2009}. Thus the time evolution can be expanded to second order as
\begin{equation}\label{expandU}
\thinmuskip=0mu
\begin{aligned}
\hat{\mathcal{U}}(t'',t')&=\mathcal{T}e^{-i\int_{t'}^{t''}[\mathcal{H}(t)+\mathcal{H}_{\textrm{probe}}(t)]dt}\\
&\approx \mathcal{U}(t'',t)-i\int_{t'}^{t''}dt\ \mathcal{U}(t'',t)\mathcal{H}_{\textrm{probe}}(t)\mathcal{U}(t,t')\\
&+{\int_{t'}^{t''}}dt_2{\int_{t'}^{t_2}}dt_1 \mathcal{U}(t'',t_2)\mathcal{H}_{\textrm{probe}}(t_2)\mathcal{U}(t_2,t_1)\\
&\times\mathcal{H}_{\textrm{probe}}(t_1)\mathcal{U}(t_1,t')
\end{aligned}
\end{equation}

The photon flux of a certain momentum and polarization, which is directly linked to spectroscopic experiments, may be measured by
\begin{equation}\label{observables}
J_{\bm{q}_{\rm f}\bm{e}_{\rm f}} =\lim_{t''\rightarrow \infty}\langle \hat{\mathcal{U}}(-\infty,t'') a^\dagger_{\bm{q}_{\rm f}\bm{e}_{\rm f}}a_{\bm{q}_{\rm f}\bm{e}_{\rm f}}\hat{\mathcal{U}}(t'',-\infty) \rangle
\end{equation}
where $t''\rightarrow \infty$ is taken to include all scattering photons, and $\langle O\rangle$ is the ensemble average of the operator $O$. Here apparently the zeroth order value is $J_{\bm{q}_{\rm f}\bm{e}_{\rm f}}^{(0)}=\delta_{\bm{q}_{\rm i} \bm{q}_{\rm f}}(\bm{e}_{\rm f}\cdot\bm{e}_{\rm i})^2$, corresponding to elastic reflection in experiments. 

In order to calculate higher-order results, first we can perform the decomposition $\mathcal{H}_{\textrm{probe}}(t)=\mathcal{H}_{\textrm{in}}(t)+\mathcal{H}_{\textrm{out}}(t)$. $\mathcal{H}_{\textrm{in}}(t)=s(t)\sum_{m\bm{k}\bm{e}}\mathcal{D}_{m\bm{e}}(t)e^{i\bm{k}\cdot\bm{r}_m}a_{\bm{k}\bm{e}}$ represents the photon absorption part with $a_{\bm{k}\bm{e}}$ annihilating a photon with momentum $\bm{k}$ and polarization $\bm{e}$, and $\mathcal{H}_{\textrm{out}}(t)=\mathcal{H}_{\textrm{in}}^\dagger(t)$. $s(t)$ is the probe envelope function, which semi-classically describes the portion of the probe field that interacts with the system. An alternative representation is keeping the interaction fixed but allowing the photon field to vary in time. These two representations lead to the same cross section obtained by the linear response theory.

\subsubsection{1st order contribution: time-resolved XAS}
The first-order contribution of $\mathcal{H}_{\textrm{probe}}(t)$ is
\begin{widetext}
\begin{equation}\label{1st-order}
J_{\bm{q}_{\rm f}\bm{e}_{\rm f}}^{(1)}={\int_{-\infty}^{\infty}}dt_2{\int_{-\infty}^{\infty}}dt_1 \langle \mathcal{U}(-\infty,t_2)\mathcal{H}_{\textrm{probe}}(t_2)\mathcal{U}(t_2,\infty) (a^\dagger_{\bm{q}_{\rm f}\bm{e}_{\rm f}}a_{\bm{q}_{\rm f}\bm{e}_{\rm f}}-J_{\bm{q}_{\rm f}\bm{e}_{\rm f}}^{(0)})\mathcal{U}(\infty,t_1)\mathcal{H}_{\textrm{probe}}(t_1)\mathcal{U}(t_1,-\infty)\rangle
\end{equation}
Here, since $\mathcal{H}_{\textrm{in}}$ corresponds to photon absorption and $\mathcal{H}_{\textrm{out}}$ corresponds to photon emission, $J_{\bm{q}_{\rm f}\bm{e}_{\rm f}}^{(1)}$ can further be decompose to two parts, i.e., $J_{\bm{q}_{\rm f}\bm{e}_{\rm f}}^{(1)}=J_{\bm{q}_{\rm f}\bm{e}_{\rm f}}^{\textrm{ab}}+J_{\bm{q}_{\rm f}\bm{e}_{\rm f}}^{\textrm{em}}$. 

The absorption part $J_{\bm{q}_{\rm f}\bm{e}_{\rm f}}^{\textrm{ab}}$ is
\begin{equation}\label{absorption}
\begin{aligned}
J_{\bm{q}_{\rm f}\bm{e}_{\rm f}}^{\textrm{ab}}&={\int_{-\infty}^{\infty}}dt_2{\int_{-\infty}^{\infty}}dt_1 \langle \mathcal{U}(-\infty,t_2)\mathcal{H}_{\textrm{out}}(t_2)\mathcal{U}(t_2,\infty) (a^\dagger_{\bm{q}_{\rm f}\bm{e}_{\rm f}}a_{\bm{q}_{\rm f}\bm{e}_{\rm f}}-J_{\bm{q}_{\rm f}\bm{e}_{\rm f}}^{(0)})\mathcal{U}(\infty,t_1)\mathcal{H}_{\textrm{in}}(t_1)\mathcal{U}(t_1,-\infty)\rangle\\
&=\sum_{\substack{m,n\\\bm{k},\bm{k'},\bm{e},\bm{e}'}}{\int_{-\infty}^{\infty}}dt_2{\int_{-\infty}^{\infty}}dt_1 s(t_1)s(t_2)\langle \mathcal{U}(-\infty,t_2)\mathcal{D}_{n\bm{e}'}^\dagger(t_2) e^{-i\bm{k}'\cdot\bm{r}_n}\mathcal{U}(t_2,t_1)\mathcal{D}_{m\bm{e}}(t_1)e^{i\bm{k}\cdot\bm{r}_m}\mathcal{U}(t_1,-\infty)\rangle\\
&~~~~~~~~~~~~~\times\big(\langle a_{\bm{k}'\bm{e}'}^\dagger(t_2) a^\dagger_{\bm{q}_{\rm f}\bm{e}_{\rm f}}a_{\bm{q}_{\rm f}\bm{e}_{\rm f}}a_{\bm{k}\bm{e}}(t_1)\rangle_{\textrm{ph}}-\langle a_{\bm{k}'\bm{e}'}^\dagger(t_2) a_{\bm{k}\bm{e}}(t_1)\rangle_{\textrm{ph}} J_{\bm{q}_{\rm f}\bm{e}_{\rm f}}^{(0)}\big),
\end{aligned}
\end{equation}
where $a_{\bm{k}\bm{e}}(t_1)=\mathcal{U}(-\infty,t_1)a_{\bm{k}\bm{e}}\mathcal{U}(t_1,-\infty)$, and $\langle O\rangle_{\textrm{ph}}$ calculates the mean value of the operator $O$ in the one-photon eigenstate $a^\dagger_{\bm{q}_{\rm i}\bm{e}_{\rm i}}\ket{0}$, which is the initial photon state in the discussed x-ray scattering process. If we have an x-ray pump, the electrons and x-ray photons will be entangled in the time evolution. The inner shell electrons will be excited to upper levels by the pump and the separation of the two parts in Eq.\eqref{absorption} will need to be carefully modified. Here we have assumed that the pump does not affect inner shells, because a typical tr-RIXS measurement takes an off-resonant pump with respect to the x-ray edges. Thus $\langle a_{\bm{k}'\bm{e}'}^\dagger(t_2) a^\dagger_{\bm{q}_{\rm f}\bm{e}_{\rm f}}a_{\bm{q}_{\rm f}\bm{e}_{\rm f}}a_{\bm{k}\bm{e}}(t_1)\rangle_{\textrm{ph}}=0$ and $\langle a_{\bm{k}'\bm{e}'}^\dagger(t_2) a_{\bm{k}\bm{e}}(t_1)\rangle_{\textrm{ph}} =e^{i\omega_{\rm i} (t_2-t_1)}\delta_{\bm{k}'\bm{q}_{\rm i}}\delta_{\bm{e}'\bm{e}_{\rm i}}\delta_{\bm{k}\bm{q}_{\rm i}}\delta_{\bm{e}\bm{e}_{\rm i}}$. 
This means we obtain no first-order contribution unless we are probing at the incident photon momentum. Then Eq.\eqref{absorption} can be simplified as 
\begin{equation}\label{xas}
\thinmuskip=0mu
\medmuskip=-1mu
\begin{aligned}
J_{\bm{q}_{\rm f}\bm{e}_{\rm f}}^{\textrm{ab}}&=-\sum_{n,m}{\int_{-\infty}^{\infty}}dt_2{\int_{-\infty}^{\infty}}dt_1 s(t_1)s(t_2)e^{i\omega_{\rm i} (t_2-t_1)}e^{i\bm{q}_{\rm i}\cdot(\bm{r}_m-\bm{r}_n)}\langle \mathcal{U}(-\infty,t_2)\mathcal{D}_{n\bm{e}_{\rm i}}^\dagger(t_2) \mathcal{U}(t_2,t_1)\mathcal{D}_{m\bm{e}_{\rm i}}(t_1)\mathcal{U}(t_1,-\infty)\rangle\delta_{\bm{q}_{\rm i}\bm{q}_{\rm f}}(\bm{e}_{\rm f}\cdot\bm{e}_{\rm i})^2\\
&=-\sum_{n}{\int_{-\infty}^{\infty}}dt_2{\int_{-\infty}^{\infty}}dt_1 s(t_1)s(t_2)e^{i\omega_{\rm i} (t_2-t_1)}\langle \mathcal{U}(-\infty,t_2)\mathcal{D}_{n\bm{e}_{\rm i}}^\dagger(t_2) \mathcal{U}(t_2,t_1)\mathcal{D}_{n\bm{e}_{\rm i}}(t_1)\mathcal{U}(t_1,-\infty)\rangle\delta_{\bm{q}_{\rm i}\bm{q}_{\rm f}}(\bm{e}_{\rm f}\cdot\bm{e}_{\rm i})^2
\end{aligned}
\end{equation}
Equation \eqref{xas} describes time-resolved x-ray absorption spectroscopy (tr-XAS). It is the first order \textit{loss} brought by the light-matter interaction at the incident energy, which corresponds to absorption of the photon by the electrons in the materials. The function $s(t_1)$ is given by the probe profile $s(t_1)=g(t_1,t)$, where $t$ is the center of the probe pulse, and the same for $s(t_2)$. 

The emission part $J_{\bm{q}_{\rm f}\bm{e}_{\rm f}}^{\textrm{em}}$ is
\begin{equation}\label{emission}
\begin{aligned}
J_{\bm{q}_{\rm f}\bm{e}_{\rm f}}^{\textrm{em}}&={\int_{-\infty}^{\infty}}dt_2{\int_{-\infty}^{\infty}}dt_1 \langle \mathcal{U}(-\infty,t_2)\mathcal{H}_{\textrm{in}}(t_2)\mathcal{U}(t_2,\infty) (a^\dagger_{\bm{q}_{\rm f}\bm{e}_{\rm f}}a_{\bm{q}_{\rm f}\bm{e}_{\rm f}}-J_{\bm{q}_{\rm f}\bm{e}_{\rm f}}^{(0)})\mathcal{U}(\infty,t_1)\mathcal{H}_{\textrm{out}}(t_1)\mathcal{U}(t_1,-\infty)\rangle\\
&=\sum_{\substack{m,n\\\bm{k},\bm{k'},\bm{e},\bm{e}'}}{\int_{-\infty}^{\infty}}dt_2{\int_{-\infty}^{\infty}}dt_1 s(t_1)s(t_2)\langle \mathcal{U}(-\infty,t_2)\mathcal{D}_{n\bm{e}'}(t_2) e^{i\bm{k}'\cdot\bm{r}_n}\mathcal{U}(t_2,t_1)\mathcal{D}_{m\bm{e}}^\dagger(t_1)e^{-i\bm{k}\cdot\bm{r}_m}\mathcal{U}(t_1,-\infty)\rangle\\
&~~~~~~~~~~~~~\times\big(\langle a_{\bm{k}'\bm{e}'}(t_2) a^\dagger_{\bm{q}_{\rm f}\bm{e}_{\rm f}}a_{\bm{q}_{\rm f}\bm{e}_{\rm f}}a_{\bm{k}\bm{e}}^\dagger(t_1)\rangle_{\textrm{ph}}-\langle a_{\bm{k}'\bm{e}'}(t_2) a_{\bm{k}\bm{e}}^\dagger(t_1)\rangle_{\textrm{ph}} J_{\bm{q}_{\rm f}\bm{e}_{\rm f}}^{(0)}\big),
\end{aligned}
\end{equation}
Equation \eqref{emission} corresponds to x-ray emission, and is non-zero only when the initial state has core-holes. For an off-resonant pump pulse that does not excite core electrons to valence levels, the contribution of Eq.\eqref{emission} can be ignored.

\subsubsection{2nd order contribution: time-resolved RIXS}
In contrast to XAS, RIXS is a photon scattering procedure and requires the participation of both incoming and outgoing photons. The second-order contribution to $J_{\bm{q}_{\rm f}\bm{e}_{\rm f}}$ can be realized by the second order scattering terms in $\mathcal{H}_{\textrm{probe}}(t)$, or a two-time sequence of the first order $\mathcal{H}_{\textrm{probe}}(t)$ in Eq.\eqref{expandU}. Resonant or non-resonant contributions can be selected by different incident photon probe energies. Here we do not consider non-resonant processes \citep{wang2018theory,freericks2018theory} but focus only on those responsible for RIXS. Such a scattering procedure can be measured by:
\begin{equation}\label{4th-order}
\thinmuskip=0mu
\medmuskip=-1mu
\begin{aligned}
&J_{\bm{q}_{\rm f}\bm{e}_{\rm f}}^{(2)\textrm{RIXS}}\\
=&{\int_{-\infty}^{\infty}} dt_2 {\int_{-\infty}^{t_2}} dt_1 {\int_{-\infty}^{\infty}} dt'_2 {\int_{-\infty}^{t'_2}} dt'_1 \langle \mathcal{U}(-\infty,t'_1)\mathcal{H}_{\textrm{out}}(t'_1)\mathcal{U}(t'_1,t'_2)\mathcal{H}_{\textrm{in}}(t'_2)\mathcal{U}(t'_2,\infty) a^\dagger_{\bm{q}_{\rm f}\bm{e}_{\rm f}}a_{\bm{q}_{\rm f}\bm{e}_{\rm f}}\mathcal{U}(\infty,t_2)\mathcal{H}_{\textrm{out}}(t_2)\mathcal{U}(t_2,t_1)\\
&\times \mathcal{H}_{\textrm{in}}(t_1)\mathcal{U}(t_1,-\infty)\rangle\\
=&\sum_{\substack{m,m'\\n,n'}}{\int_{-\infty}^{\infty}} dt_2 {\int_{-\infty}^{t_2}} dt_1 {\int_{-\infty}^{\infty}} dt'_2 {\int_{-\infty}^{t'_2}} dt'_1 \langle \mathcal{U}(-\infty,t'_1)\mathcal{D}_{n'\bm{\varepsilon}'}^\dagger(t'_1) \mathcal{U}(t'_1,t'_2)\mathcal{D}_{n\bm{\varepsilon}}(t'_2)\mathcal{U}(t'_2,t_2)\mathcal{D}_{m'\bm{e}'}^\dagger(t_2) \mathcal{U}(t_2,t_1)\mathcal{D}_{m\bm{e}}(t_1) \mathcal{U}(t_1,-\infty)\rangle\\
&\times s(t'_1)s(t'_2)s(t_2)s(t_1) \sum_{\substack{\bm{k_1},\bm{k_2},\bm{k'_2},\bm{k'_1}\\ \bm{e},\bm{e}',\bm{\varepsilon},\bm{\varepsilon}'}}e^{i(\bm{k}_1\cdot \bm{r}_m - \bm{k}_2\cdot \bm{r}_{m'}+\bm{k}'_2\cdot \bm{r}_n-\bm{k}'_1\cdot \bm{r}_{n'})}\langle a_{\bm{k}'_1\bm{\varepsilon}'}^\dagger(t'_1) a_{\bm{k}'_2\bm{\varepsilon}}(t'_2)a^\dagger_{\bm{q}_{\rm f}\bm{e}_{\rm f}}a_{\bm{q}_{\rm f}\bm{e}_{\rm f}}a_{\bm{k}_2\bm{e}'}^\dagger(t_2)a_{\bm{k}_1\bm{e}}(t_1)\rangle_{\textrm{ph}}\\
=&\sum_{m,n}{\int_{-\infty}^{\infty}} dt_2 {\int_{-\infty}^{t_2}} dt_1 {\int_{-\infty}^{\infty}} dt'_2 {\int_{-\infty}^{t'_2}} dt'_1 \langle \mathcal{U}(-\infty,t'_1)\mathcal{D}_{n\bm{\varepsilon}'}^\dagger(t'_1) \mathcal{U}(t'_1,t'_2)\mathcal{D}_{n\bm{\varepsilon}}(t'_2)\mathcal{U}(t'_2,t_2)\mathcal{D}_{m\bm{e}'}^\dagger(t_2) \mathcal{U}(t_2,t_1)\mathcal{D}_{m\bm{e}}(t_1) \mathcal{U}(t_1,-\infty)\rangle\\
&\times s(t'_1)s(t'_2)s(t_2)s(t_1)\sum_{\substack{\bm{k_1},\bm{k_2},\bm{k'_2},\bm{k'_1}\\ \bm{e},\bm{e}',\bm{\varepsilon},\bm{\varepsilon}'}} e^{i(\bm{k}_1 - \bm{k}_2)\cdot \bm{r}_{m}+i(\bm{k}'_2-\bm{k}'_1)\cdot \bm{r}_{n}}\langle a_{\bm{k}'_1\bm{\varepsilon}'}^\dagger(t'_1) a_{\bm{k}'_2\bm{\varepsilon}}(t'_2)a^\dagger_{\bm{q}_{\rm f}\bm{e}_{\rm f}}a_{\bm{q}_{\rm f}\bm{e}_{\rm f}}a_{\bm{k}_2\bm{e}'}^\dagger(t_2)a_{\bm{k}_1\bm{e}}(t_1)\rangle_{\textrm{ph}}.
\end{aligned}
\end{equation}
\end{widetext}
For tr-RIXS, $\bm{q}_{\rm f}=\bm{q}_{\rm i}-\bm{q}$, and the initial photon state is $a^\dagger_{\bm{q}_{\rm i}\bm{e}_{\rm i}}\ket{0}$. We may evaluate the photon part in Eq.\eqref{4th-order} to be $\langle a_{\bm{k}'_1\bm{\varepsilon}'}^\dagger(t'_1) a_{\bm{k}'_2\bm{\varepsilon}}(t'_2)a^\dagger_{\bm{q}_{\rm f}\bm{e}_{\rm f}}a_{\bm{q}_{\rm f}\bm{e}_{\rm f}}a_{\bm{k}_2\bm{e}'}^\dagger(t_2)a_{\bm{k}_1\bm{e}}(t_1)\rangle_{\textrm{ph}}=\delta_{\bm{k}_1\bm{q}_{\rm i}}\delta_{\bm{k}'_1\bm{q}_{\rm i}}\delta_{\bm{k}_2\bm{q}_{\rm f}}\delta_{\bm{k}'_2\bm{q}_{\rm f}}e^{i\omega_{\rm i}(t'_1-t_1)-i\omega_{\rm f}(t'_2-t_2)}\delta_{\bm{e}\bm{e}_{\rm i}}\delta_{\bm{\varepsilon}'\bm{e}_{\rm i}}\delta_{\bm{e}'\bm{e}_{\rm f}}\delta_{\bm{\varepsilon}\bm{e}_{\rm f}}$. Note that $t_1$ and $t'_1$ correspond to the excitation process, while $t_2$ and $t'_2$ correspond to the de-excitation process. Equation \eqref{4th-order} provides the scattered photon flux from the many-body system, which evolves first up to time $t_1$, when an electron from the core level is resonantly excited to the valence band, and then further evolves the many-body state (with the core hole) up to time $t_2$. At this point, the valence electron drops back to eliminate the core hole. Such evolution is represented by the Keldysh contour shown in Fig.\ref{fig:1}. 

Typically, we do not consider the phenomenological lifetime of the excitations or quasiparticles, since it is usually longer than the probe pulse for low-energy excitations. However, since the core hole has a huge binding energy, its lifetime has to be explicitly considered. This means the refilling of the core hole at $t_2$ must happen within a certain time window after it was created at $t_1$. The detailed evaluation of this procedure requires the consideration of the interaction with the environment. Tracing out the environmental degrees of freedom then leads to a non-Hermitian decay process, whose net effect is a non-unitary $\mathcal{U}(t_2,t_1)$. As a phenomenological description of the irreversible decay process, we modify Eq.\eqref{expandU} and Eq.\eqref{4th-order} for the time intervals $[t_1,t_2]$  and $[t'_1,t'_2]$, where the core hole exists, with $\mathcal{U}(t_2,t_1)\rightarrow l(t_1,t_2)\mathcal{U}(t_2,t_1)$ and $\mathcal{U}(t'_1,t'_2)\rightarrow l(t'_2,t'_1)\mathcal{U}(t'_1,t'_2)$, where $l(t_1,t_2)=\exp(-|t_2-t_1|/\tau_{\textrm{ch}})$ describes the lifetime of the core hole. The probe pulse is treated the same as in XAS described above, i.e., $s(\tau)=g(\tau,t)$, where $t$ is the observation time. To simply the expression, we also define the four-point correlation function
\begin{equation}
\thinmuskip=0mu
\medmuskip=-1mu
\begin{aligned}
S_{\bm{e}_{\rm i} \bm{e}_{\rm f}}^{mn}(t_1,&t_2,t'_2,t'_1)=\langle U(-\infty,t'_1)\mathcal{D}_{n\bm{e}_{\rm i}}^\dagger(t'_1) U(t'_1,t'_2)\mathcal{D}_{n\bm{e}_{\rm f}}(t'_2)\\
&\times U(t'_2,t_2)\mathcal{D}_{m\bm{e}_{\rm f}}^\dagger(t_2) U(t_2,t_1)\mathcal{D}_{m\bm{e}_{\rm i}}(t_1) U(t_1,-\infty)\rangle
\end{aligned}
\end{equation}
Putting all of this together, we find
\begin{align}\label{trRIXS}
&I(\omega_{\rm i},\omega_{\rm f},\bm{q},t)=J_{\bm{q}_{\rm f}\bm{e}_{\rm f}}^{(2)\textrm{RIXS}}\nonumber\\
=&{\int_{-\infty}^{\infty}} dt_2 {\int_{-\infty}^{t_2}} dt_1 {\int_{-\infty}^{\infty}} dt'_2 {\int_{-\infty}^{t'_2}} dt'_1 e^{i\omega_{\rm i}(t'_1-t_1)-i\omega_{\rm f}(t'_2-t_2)}\nonumber\\
&\times l(t_1,t_2)l(t'_1,t'_2)g(t_1,t)g(t_2,t)g(t'_1,t)g(t'_2,t)\nonumber\\
&\times \sum_{m,n}e^{i\bm{q}\cdot (\bm{r}_{m}-\bm{r}_{n})}S_{\bm{e}_{\rm i} \bm{e}_{\rm f}}^{mn}(t_1,t_2,t'_2,t'_1)\nonumber\\
\end{align}
Equation \eqref{trRIXS} is the full cross section of tr-RIXS. The dipole excitation and dipole de-excitation processes form the scattering amplitude, resulting in a closed-form cross section determined by $t_1<t_2$, $t'_1<t'_2$. We should note that the derivation of the tr-RIXS cross section here is generic and not dependent on a particular form of $\mathcal{H}(t)$ or a specific probe shape $s(t)$. 
\subsection{Comparison to Equilibrium Kramers-Heisenberg Formula}\label{sec:equilibrium}
As a special case of tr-RIXS, the equilibrium RIXS cross section can be obtained by assuming a time-independent Hamiltonian, i.e., $\mathcal{H}(t)=\mathcal{H}(t+\tau)=\mathcal{H}$ for any $\tau$. Assuming zero temperature, we evaluate the four-point correlation function to be
\begin{equation}
\begin{aligned}
S_{\bm{e}_{\rm i} \bm{e}_{\rm f}}^{mn}&(t_1,t_2,t'_2,t'_1)=\bra{\Psi_{\rm i}^0}\mathcal{D}_{n\bm{e}_{\rm i}}^\dagger e^{-i(\mathcal{H}-E_{\rm i})(t'_1-t'_2)} \mathcal{D}_{n\bm{e}_{\rm f}}\\
&\times e^{-i(\mathcal{H}-E_{\rm i})(t'_2-t_2)}\mathcal{D}_{m\bm{e}_{\rm f}}^\dagger e^{-i(\mathcal{H}-E_{\rm i})(t_2-t_1)} \mathcal{D}_{m\bm{e}_{\rm i}}\ket{\Psi_{\rm i}^0}
\end{aligned}
\end{equation}
where $\ket{\Psi_{\rm i}^0}$ is the initial ground state with energy $E_{\rm i}$. For the equilibrium case, Eq.\eqref{trRIXS} should be understood in terms of a constant rate of detection for the scattered photons\citep{FreericksPRL2009, freericks2009theoretical}, and we should set the probe shape function $g(t_i,t)=1$. This then yields
\begin{equation}\label{trRIXS_modified}
\begin{aligned}
I(\omega_{\rm i},&\omega_{\rm f},\bm{q},t)\propto\lim_{L\rightarrow\infty}\frac{1}{2L}\int_{-L}^{L}dt_2 \int_{-L}^{t_2}dt_1\int_{-L}^{L}dt'_2 \int_{-L}^{t'_2}dt'_1\\
&\times e^{i\omega_{\rm i}(t'_1-t_1)-i\omega_{\rm f}(t'_2-t_2)} l(t_1,t_2)l(t'_1,t'_2)\\
&\times\sum_{m,n}e^{i\bm{q}\cdot(\bm{r}_m-\bm{r}_n)}S_{\bm{e}_{\rm i} \bm{e}_{\rm f}}^{mn}(t_1,t_2,t'_2,t'_1)
\end{aligned}
\end{equation}
Through the transformation $r=t_2-t_1, s=t'_2-t_2, \tau=t'_2-t'_1, T=(t_1+t'_1)/2-t$, we obtain
\begin{align}\label{trRIXS_modified_app}
&I(\omega_{\rm i},\omega_{\rm f},\bm{q},t)\nonumber\\
\propto&\lim_{L\rightarrow\infty}\frac{1}{2L}\int_{0}^{2L}d\tau \int_{-2L}^{2L}ds\int_{0}^{2L}dr \int_{-L-t}^{L-t}dT\ e^{-\Gamma r}e^{-\Gamma\tau}\nonumber\\
&\times\sum_{m,n}e^{i\bm{q}\cdot(\bm{r}_m-\bm{r}_n)}\bra{\Psi_{\rm i}^0}\mathcal{D}_{n\bm{e}_{\rm i}}^\dagger e^{-i(\omega_{\rm i}-\mathcal{H}+E_{\rm i})\tau} \mathcal{D}_{n\bm{e}_{\rm f}}\nonumber\\
&\times e^{i(\Delta\omega-\mathcal{H}+E_{\rm i})s}\mathcal{D}_{m\bm{e}_{\rm f}}^\dagger e^{i(\omega_{\rm i}-\mathcal{H}+E_{\rm i})r} \mathcal{D}_{m\bm{e}_{\rm i}}\ket{\Psi_{\rm i}^0}\nonumber\\
\propto& \lim_{L\rightarrow\infty}\int_{0}^{2L}d\tau \int_{-2L}^{2L}ds\int_{0}^{2L}dr \nonumber\\
&\sum_{m,n}\bra{\Psi_{\rm i}^0}\mathcal{D}_{n\bm{e}_{\rm i}}^\dagger e^{-\Gamma\tau-i(\omega_{\rm i}-\mathcal{H}+E_{\rm i})\tau} \mathcal{D}_{n\bm{e}_{\rm f}}e^{-i\bm{q}\cdot\bm{r}_n}e^{i(\Delta\omega-\mathcal{H}+E_{\rm i})s}\nonumber\\
&\times e^{i\bm{q}\cdot\bm{r}_m}\mathcal{D}_{m\bm{e}_{\rm f}}^\dagger e^{-\Gamma r+i(\omega_{\rm i}-\mathcal{H}+E_{\rm i})r} \mathcal{D}_{m\bm{e}_{\rm i}}\ket{\Psi_{\rm i}^0}
\end{align}
where the inverse core-hole lifetime $\Gamma = 1/\tau_{\textrm{ch}}$. Using $\int_{0}^{\infty}dr \ e^{-\delta r+i\alpha r}=i(\alpha+i\delta)^{-1}$ and $\int_{-\infty}^{\infty}e^{ics}ds=2\pi\delta(c)$ (for real $c$), we obtain the equilibrium RIXS cross-section
\begin{equation}\label{compare_KH}
I_{\rm eq}(\omega_{\rm i},\omega_{\rm f},\bm{q},t)=\sum_{\rm f}|A_{\rm f}|^2 \delta(\Delta\omega-E_{\rm f}+E_{\rm i})
\end{equation}
where $\rm f$ labels an eigenstate $\ket{\Psi_{\rm f}^0}$ of $\mathcal{H}$ with energy $E_{\rm f}$, and
\begin{equation}\label{A}
A_{\rm f}=\bra{\Psi_{\rm f}^0}\sum_{m}e^{i\bm{q}\cdot\bm{r}_m}\mathcal{D}_{m\bm{e}_{\rm f}}^\dagger\frac{1}{\omega_{\rm i}-\mathcal{H}+E_{\rm i}+i\Gamma}\mathcal{D}_{m\bm{e}_{\rm i}}\ket{\Psi_{\rm i}^0}
\end{equation}
is the scattering amplitude. This recovers the Kramers-Heisenberg formula for equilibrium RIXS\cite{TPDReview,ChunjingPRX,DemlerPRL2014}.

The above derivation assumes infinite probe width. In practice, a finite probe width typically gives a linewidth on top of the $\delta$-functions in photoemission or Raman scattering\cite{FreericksPRL2009,wang2018theory}. Similarly in RIXS, the conventional Kramers-Heisenberg formula Eq.\eqref{KS} differs from Eq.\eqref{compare_KH} by a consideration of finite linewidth as a result of the probe shape. However, we notice that this linewidth is in fact a complicated form instead of an inverse of the probe width. That is because in Eq.\eqref{trRIXS_modified_app} the part that corresponds to energy loss $\Delta\omega$ only contains the photon emission time $t_2$ and $t_2'$, while the probe shape contains $t_1$ and $t_1^\prime$. The consequence is that the final effective linewidth, if written in the Kramers-Heisenberg form, contains a renormalization by the core-hole lifetime $\tau_{\rm ch}$. Conceptually it indicates the ``excitations'' effectively develops at a weighted average time between $t_1$ and $t_2$ (or $t_1^\prime$ and $t_2^\prime$). When the core-hole lifetime is much shorter than the probe duration, which is typical in experiments, we can simply reduce the shape function by $g(t_1,t)g(t_2,t)g(t'_1,t)g(t'_2,t)\approx g^2(t_2,t)g^2(t'_2,t)$ similar to the nonresonant scattering\cite{wang2018theory,freericks2018theory}. This recovers the well-known Kramers-Heisenberg formula. 
\begin{figure*}[!ht]
\begin{center}
\includegraphics[width=18cm]{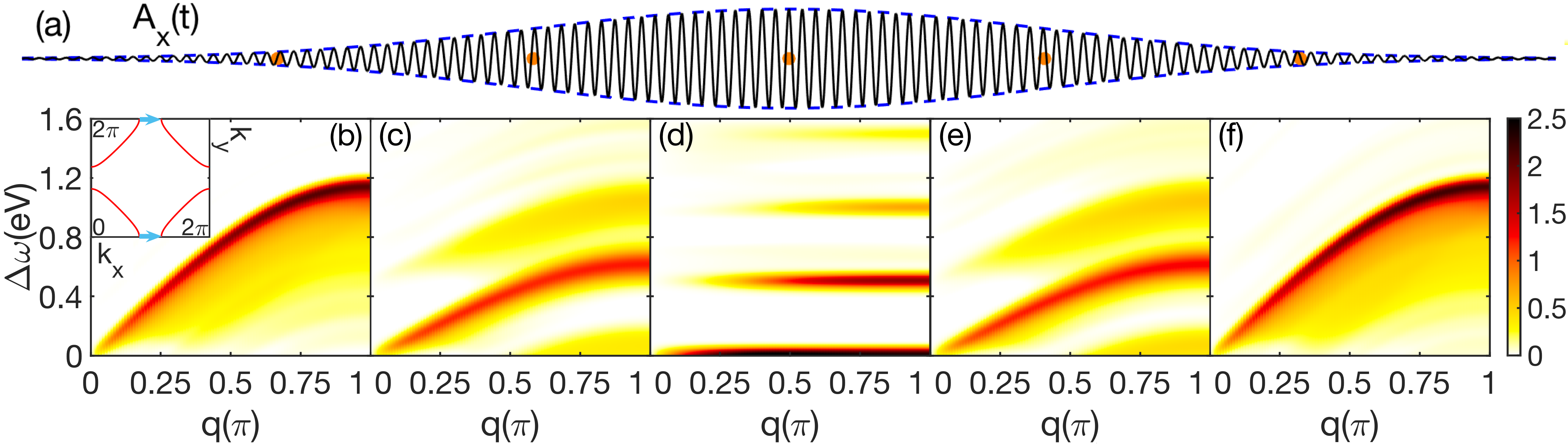}
\caption{\label{fig:2} Tr-RIXS spectra for a large pump frequency. (a) Schematic of the pump pulse. The Gaussian profile is shown in blue dashed lines, and the five orange dots correspond to the times when spectra in (b)-(f) are taken. (b)-(f) The tr-RIXS spectra at $t=-2\sigma_{\textrm{pu}},-\sigma_{\textrm{pu}},0,\sigma_{\textrm{pu}},2\sigma_{\textrm{pu}}$, respectively. The inset of (b) shows the Fermi surface of the 2D electronic system with red lines, and the blue arrows show the nesting momentum wavevector at $q\simeq 0.4\pi$. Unless otherwise specified, all RIXS intensities are shown in the same arbitrary units.
}
\end{center}
\end{figure*}
\section{Numerical calculation of tr-RIXS}
\label{num}
In order to provide a basic picture of the nonequilibrium physics revealed from tr-RIXS, we study electrons on a 2D square lattice at zero temperature as a simple example. Although an explicit treatment of the electron interaction is necessary in quantum materials with strong correlations, a precise description of the many-body states is then necessarily restricted to a small cluster, which conceals the momentum resolution of RIXS. As a benchmark study focused on the tr-RIXS measurements, the conduction band can be written in a tight-binding form:
\begin{equation}\label{tight_binding}
\varepsilon_{\bm{k}}=\varepsilon_c+\mu_0-2t_h\big(\cos k_x+\cos k_y\big)
\end{equation}
in which $\varepsilon_c$ is the energy of the core level, $\mu_0$ is the energy difference between core and valence electrons in the atomic limit, and $t_h$ is the hopping of the conduction band which is chosen to be 0.3eV. The symbol $\varepsilon_{\bm{k}}$ denotes the band dispersion. When there is an external uniform pump field $\bm{A}(t)$, the Peierl's substitution gives $\bm{k}\rightarrow\bm{k}-\bm{A}(t)$ in Eq.\eqref{tight_binding}\citep{peierls1933}. 
The probe energy is set to be $\omega_{\rm i}=\mu_0$ to achieve resonance. We can explicitly evaluate Eq.\eqref{trRIXS} to find
\begin{widetext}
\begin{equation}\label{BQ}
\thinmuskip=0mu
\medmuskip=-1mu
\begin{aligned}
I(&\Delta\omega, \bm{q},t)\propto\iiiint_{t_1<t_2,t'_1<t'_2}dt_1dt_2dt'_2dt'_1\  e^{i(\omega_{\rm i}-\mu_0) (t'_1-t_1)-i(\omega_{\rm i}-\Delta\omega-\mu_0) (t'_2-t_2)}g(t_1,t)g(t_2,t)g(t'_1,t)g(t'_2,t)l(t_1,t_2)l(t'_1,t'_2)\\
&\times\sum_{\bm{k}}\bigg\{ f(\varepsilon_{\bm{k}})\big[1-f(\varepsilon_{\bm{k}+\bm{q}})\big]e^{-i\int_{t_2}^{t'_2} d\tau (2t_h)\left[\text{cos}\big(k_x - A_x(\tau)\big)+ \text{cos}\big(k_y - A_y(\tau)\big)\right]+i\int_{t_1}^{t'_1} d\tau (2t_h)\left[\text{cos}\big(k_x+q_x - A_x(\tau)\big)+ \text{cos}\big(k_y+q_y - A_y(\tau)\big)\right]}\bigg\}\\
\end{aligned}
\end{equation}
\end{widetext}
where $f(\varepsilon)$ is the Fermi-Dirac distribution. 
To mimic the probe profile and the phenomenological decay, we employ $g(\tau,t)=1/(\sqrt{2\pi}\sigma_{\textrm{pr}})\exp[-(\tau-t)^2/2\sigma_{\textrm{pr}}^2]$ and $l(\tau,t)=\exp[-|\tau-t|/\tau_{\textrm{ch}}]$ in Eq.\eqref{BQ}. Note that we do not incorporate any specific time-dependent form for the dipole matrix elements, nor do we correct the momentum shift as is done in tr-ARPES to produce a gauge-invariant Green's function\citep{freericks2015gauge,bertoncini1991gauge}. The issue of gauge-invariance in the presence of a pump is a complex one, which we defer to future work. 

\subsection{High Frequency Pump: Floquet Physics}\label{high_w}
Here we choose a pump having a Gaussian profile, polarized along the Brillouin zone diagonal. The width of the pump is chosen to be $\sigma_{\textrm{pu}}=240\textrm{eV}^{-1}=151\textrm{fs}$, the maximum strength of the pump in both directions is $A_0$, and the pump frequency is $\Omega$. Here $1\textrm{eV}^{-1}=\hbar/\textrm{eV}=0.628\textrm{fs}$. Then the time-dependent pump is given by $A_x(t)=A_y(t)=A_0\exp(-t^2/2\sigma_{\textrm{pu}}^2)\cos\left(\Omega t\right)$. The width of the incoming probe pulse is $\sigma_{\textrm{pr}}=30\textrm{eV}^{-1}=18.8\textrm{fs}$. The core-hole life time is $\tau_{\textrm{ch}}=1.5\textrm{eV}^{-1}$. The choice of parameters is motivated by the hierachy of time scales in the system: the probe width determines the balance between time and energy resolution, and the pump width guarantees a nontrivial drive and relatively long steady state for given pump frequency\citep{Sentef2015,kalthoff2018emergence}. The Fermi energy of the 2D system is taken to be $\varepsilon_F=\varepsilon_c+\mu_0-0.1\textrm{eV}=\mu-0.1\textrm{eV}$, where $\mu$ is the energy of the band center. A 200$\times$200 momentum grid in the Brillouin zone is chosen in the calculation, and the time step for numerical integration is chosen to be 0.12$\textrm{eV}^{-1}$.

Figure \ref{fig:2} shows the time-resolved RIXS spectra at $t=-2\sigma_{\textrm{pu}}$, $-\sigma_{\textrm{pu}}$, $0$, $\sigma_{\textrm{pu}}$, and $2\sigma_{\textrm{pu}}$, respectively, with $\Omega=0.5$eV and $A_0= 2.4$. The transferred momentum $\bm{q}$ lies along the $x$ direction, i.e., $\bm{q}=(q,0)$. At the beginning and the end of the pump (i.e., $t=-2\sigma_{\textrm{pu}}$ and $2\sigma_{\textrm{pu}}$), the RIXS spectra are similar to the equilibrium result\citep{TPD2016}. The RIXS spectra depict the particle-hole continuum of the band: the excitation softens to zero at the (weak) nesting momentum $q\simeq 0.4\pi$ that spans the Fermi surface [as shown in Fig.\ref{fig:2}(b) inset] along $(\pi,0)$ at the given $\varepsilon_F$, and strong intensity is observed at $q=\pi, \Delta\omega\simeq1.2\textrm{eV}$, consistent with the Lindhard response (see more discussion in Appendix A).

When approaching the pump center, the spectra are flattened (i.e., become essentially dispersionless for most $q$), together with the appearance of replicas of the spectra. This phenomenon is most prominent at the pump center [see Fig.\ref{fig:2}(d)], with completely flat spectra and replicas at multiples of the pump energy. When the pump field fades away as $t>2\sigma_{\textrm{pu}}$, the spectrum recovers to the equilibrium result before the pump. Due to the lack of many-body interactions, there is no net excitation induced after the pump terminates since the electron distribution $n_{\bm{k}}$ is conserved. In addition,
at times $t$ and $-t$, the spectra are the same to numerical accuracy because the symmetric pump profile obeys time-reversal symmetry. 


\begin{figure*}[!ht]
\begin{center}
\includegraphics[width=16cm]{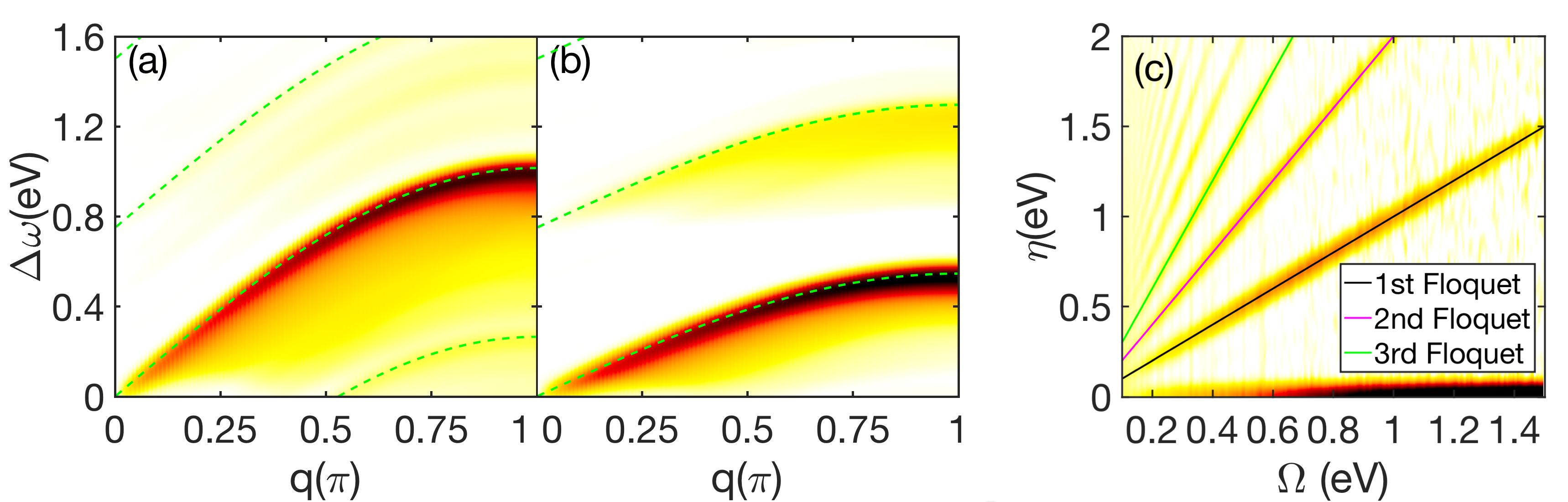}
\caption{\label{fig:3}
(a),(b) Tr-RIXS spectra for different pump amplitudes $A_0$ at the pump center: (a)$A_0=0.8$ and (b)$A_0=1.6$. Both pumps lie along $x=y$ with frequency $\Omega=0.75\textrm{eV}$. The dashed lines are the Floquet theoretical upper bounds of branches of the particle-hole continuum. The color scale is the same as in Fig.\ref{fig:2}. (c) The autocorrelation $\mathcal{C}(\bm{q},\eta,t)$ for $\bm{q}=(\pi,0)$ at $t=0$. The black, magenta, and green lines show the theoretical 1st, 2nd, and 3rd Floquet frequencies $n\Omega$, respectively. $A_0$ is kept at 2.4 throughout.
}
\end{center}
\end{figure*}
The replica features and the energy renormalization in tr-RIXS spectra can be explained in the Floquet framework. A system driven by a time-periodic external field is described by Floquet theory \citep{Floquet1883, Floquet1965}. Combined with spatial periodicity, a lattice under periodic excitation would have Floquet-Bloch electronic bands that exhibit periodicity in both momentum and energy. The steady-state Floquet theory, with an infinitely long pump, predicts two critical signatures at the lowest order: the replicas above and below the original bands separated by integer multiples of the pump frequency, and the band renormalization by $\mathcal{J}_0(A_0)$, where $\mathcal{J}_0$ is the Bessel function of the first kind \citep{Floquet1965,FloquetTheory,YaoNoneq}. The steady state band structure is renormalized into
\begin{equation}\label{Floquet_band}
E_{\bm{k},n}=\mu-2t_h\smashoperator{\sum_{\alpha=x,y}}\mathcal{J}_0(A_{\alpha 0})\cos(k_{\alpha})\pm n\Omega
\end{equation}
where $A_{\alpha 0}$ is the constant amplitude of a periodic pump field in the $\alpha$ direction, and $n$ is an integer for the replicas. More discussion on the derivation of Eq.\eqref{Floquet_band} can be found in Appendix B. Complementary to the substantially studied single-particle Floquet properties\citep{Gedik2013Science, FTI2011Galitski,FTI2013Galitski, NuhGedik2016NP}, the tr-RIXS here examines the multiparticle features associated with Floquet physics, with both energy and momentum resolution. In our calculation, $\sigma_{\textrm{pu}}$ is much larger than the pump oscillation period $2\pi/\Omega$ and the core-hole lifetime $\tau_{\textrm{ch}}$; thus the transient states can be approximated by the steady state at the (slowly varying) instantaneous pump amplitude \citep{kalthoff2018emergence}, consistent with the experimental setup in \citep{Dean2016trRIXS}. Due to the changes of the single-particle band structure, the particle-hole excitations revealed in tr-RIXS also exhibit the replicas and renormalization signatures. In this case, the transient states are manipulated adiabatically by the pump field, leaving the energy renormalization predictable by the steady-state assumption. At the pump center, the flat band arises as the pump strength $A_{\alpha0}=2.4$ is close to the zero of the Bessel function $\mathcal{J}_0$. 

To verify the above interpretation and the relation to Floquet physics, we examine the tr-RIXS for different pump amplitudes in Fig.\ref{fig:3}. For a given momentum transfer $\bm{q}=(q,0)$, let $\bm{k}(\bm{k'})$ be the momentum of the electron decaying to (excited from) the core. The upper bound of the particle-hole excitation given $q$ can be achieved at the maximum of $E_{\bm{k'},n}-E_{\bm{k},n}$. From Eq.\eqref{Floquet_band} this happens when their momenta along the $x$ direction satisfy $k_x+k_x' = \pi$. Together with $q+k_x=k'_x$, we can conclude that the upper bound of each branch of the particle-hole continuum is $\Delta\omega=4t_h\mathcal{J}_0(A_{0})\sin\frac{q}{2}\pm n\Omega$, as shown in dashed lines in Figs.\ref{fig:3}(a) and (b), agreeing well with calculated spectra. This consistency further confirms our interpretation of the multi-particle features through the steady-state assumption.

On the other hand, the steady-state assumption can also be directly validated through the change of the pump frequencies. To quantify the replica features, we define the autocorrelation via\citep{YaoNoneq}
\begin{equation}\label{autocorrelation}
\mathcal{C}(\bm{q}, \eta, t){\propto}\int_{-\infty}^{\infty}I(\omega,\bm{q},t)I(\omega+\eta,\bm{q},t) d\omega
\end{equation}
which represents the self-similarity of the nonequilibrium spectral function with a shift $\eta$ in the energy loss. In Fig.\ref{fig:3}(c), we plot the autocorrelation at $q_x=\pi$ which has the highest spectral weight at $t=0$ along $q_y=0$. The strong autocorrelation along with the theoretical Floquet predictions confirm that branches of the particle-hole excitations are separated by the pump frequency. This is a consequence of the pump-induced redistribution of the electrons to different sidebands separated by $n\Omega$ without many-body scattering. The autocorrelation map confirms that the steady-state assumption is valid and the Floquet physics dominates the dynamics for a large pump frequency.
\begin{figure*}[!ht]
\begin{center}
\includegraphics[width=18cm]{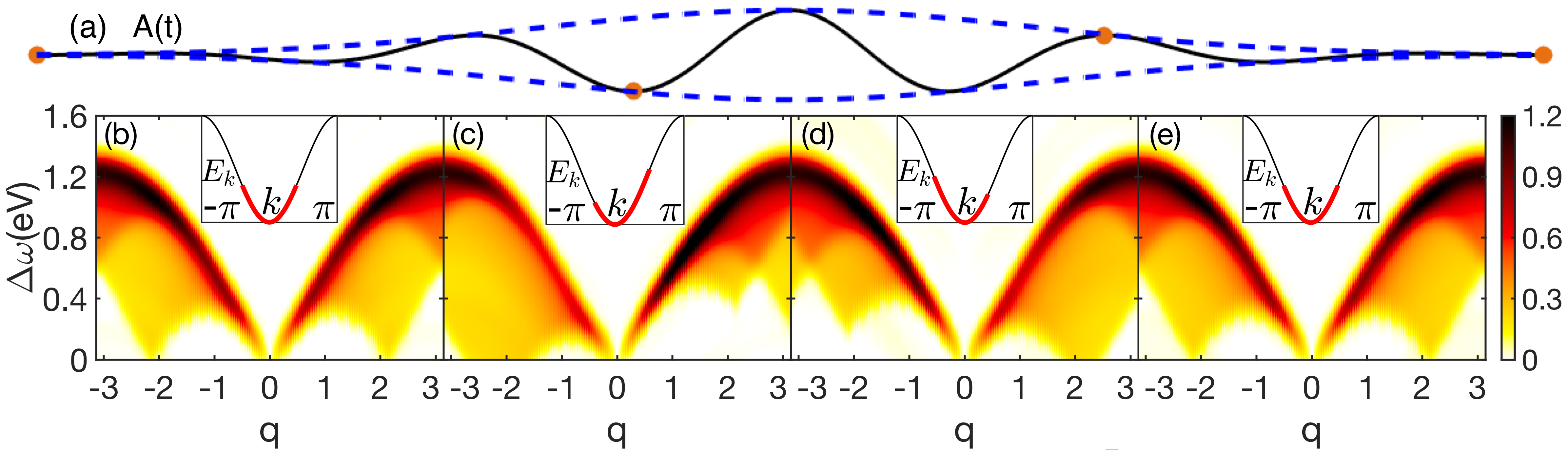}
\caption{\label{fig:4}
Tr-RIXS spectra of the 1D system with a low pump frequency. (a) Sxchematic picture of the pump pulse. The Gaussian profile is shown in blue dashed lines, and the four orange dots correspond to the time when spectra in (b)-(e) are taken. (b)-(e) The tr-RIXS spectra at $t=-720\textrm{eV}^{-1}, -150\textrm{eV}^{-1}, 300\textrm{eV}^{-1}, 720\textrm{eV}^{-1}$ respectively. The insets plot the band structure and electron distributions in momentum space (shown in red) at each time point.
}
\end{center}
\end{figure*}
\subsection{Low Frequency Pump: Effect on nesting conditions in one dimension}

The Floquet steady-state picture is valid only when the pump has high frequency with much shorter period than the duration of both the pump and probe pulse. However, this can be violated for the pump at small frequency, e.g.\ when driving the system through a phonon mode. In this case, we expect tr-RIXS to reflect adiabatic real-time dynamics of the particle-hole excitations. For a 1D electronic system, an important signature is the momentum nesting at $2k_F$ where low energy particle-hole excitations exist. It is important to see the adiabatic dynamics of such signatures under an external drive, so we calculated the tr-RIXS of a 1D tight-binding model with the same $t_h$ as before and we set $k_F=0.34\pi\approx1.07$ which means it is hole-doped. To reflect the new time scale, the pump width is still $\sigma_{\textrm{pu}}=240\textrm{eV}^{-1}=151\textrm{fs}$ but with a much lower frequency $\Omega=0.02\textrm{eV}$. It is aligned along the 1D chain with a small amplitude $A_0=0.4$. The probe width is $\sigma_{\textrm{pr}}=15\textrm{eV}^{-1}=9.4\textrm{fs}$ to balance the time and energy resolution. Note that the probe width is chosen to be much smaller than the period of the pump oscillation. The core-hole lifetime here is $\tau_{\textrm{ch}}=0.5\textrm{eV}^{-1}$, still very small compared to $\sigma_{\textrm{pr}}$.  

Figure \ref{fig:4} shows the tr-RIXS spectra of this 1D system at $t=-720\textrm{eV}^{-1}$, $-150\textrm{eV}^{-1}$, $300\textrm{eV}^{-1}$, and $720\textrm{eV}^{-1}$, respectively. At $t=-720\textrm{eV}^{-1}$, the pump has just been turned on, and we can clearly see the momentum nesting in Fig.\ref{fig:4}(b) with zero energy excitation just at $q=\pm 2k_F\approx\pm 2.14$. However, as the pump gets stronger, the momentum nesting features become different. Under the pump $\bm{A}(t)$, an electron that had momentum $\bm{k}$ initially will appear at $\bm{k}-\bm{A}(t)$ adiabatically. As the electrons oscillate in momentum space, the tr-RIXS spectrum near $2k_F$ ``breathes'' accordingly. At $t=-150\textrm{eV}^{-1}$, $A(t)$ is negative, and the electrons move to the $+x$ direction along the cosine band. This makes the excitation at $2k_F$ have a nonzero energy while a finite range of momenta around $-2k_F$ could have zero energy excitations. As observed in Fig.\ref{fig:4}(c), the nesting point at $2k_F$ lifts off and the nesting at $-2k_F$ gets flattened. Similarly, at $t=300\textrm{eV}^{-1}$, $A(t)$ is positive and the electrons move to the $-x$ direction adiabatically, as shown in the inset of Fig.\ref{fig:4}(d). The nesting at $-2k_F$ lifts off while the $2k_F$ nesting gets flattened. Finally, at $t=720\textrm{eV}^{-1}$, the pump is almost zero and the spectrum recovers to the original form. 
Since tr-RIXS can be used to track momentum nesting, it would be interesting to see how the charge density waves change as well when the pump is on. It has been theoretically proposed that most charge density waves actually result from strong electron-phonon interaction rather than Fermi surface nesting\citep{johannes2008fermi}. Therefore, by tracking momentum nesting and charge density waves at the same time in tr-RIXS experiments, we may answer whether Fermi surface nesting is the origin of charge density waves in certain materials.

\section{Discussion and Conclusion}\label{diss}
The RIXS process studied here is direct RIXS: the incoming photon promotes a core electron to an empty valence band state, and then an electron from a different state in the valence band decays and annihilates the core hole\citep{TPDReview}. The core hole itself does not leave quasiparticles after it is filled. In contrast to the crucial role core-hole attraction plays in indirect RIXS where it scatters valence electrons and creates excitations in the intermediate state\citep{TPDReview,ChunjingNJP,Jia2014NComm,tohyama2018spectral}, for direct RIXS it does not manifest explicitly. Currently we have not included the core-hole potential in this work, since it is computationally expensive. In our future work, we will include the core-hole potential for systems in which it is important.

Tr-RIXS has opened a gate for investigating a variety of important physical problems. A pump can be used to achieve band engineering. Tr-RIXS can detect momentum nesting and charge density waves changed by such band engineering and shed light on their relationship. Tr-RIXS can also decipher the topological effects in charge excitations brought by the pump in materials such as transition-metal dichalcogenides \citep{claassen2016all}. Besides, in systems with spin-orbit coupling, we can use tr-RIXS to track the interplay of charge and spin excitations under an external pump\citep{wang2014real}. Our theory does not depend on the form of the Hamiltonian and thus could be used to study materials with strong correlation. 

In conclusion, this study proposed a generic theoretical formalism for calculating time-resolved resonant spectroscopies, including tr-XAS and tr-RIXS. We benchmarked this method on a non-interacting system and calculated the tr-RIXS spectra under a pump pulse. The evolution of the particle-hole excitations can be resolved in the spectra. For a high-frequency pump, the tr-RIXS spectrum displays both replica excitations and band renormalization. Through an autocorrelation analysis for various pump frequencies, we found the features can be captured by a steady-state assumption and Floquet theory for multi-particle excitations. However, when the pump frequency is low, the dynamics of the system behaves more adiabatically and displays a breathing of the spectrum near the nesting momentum, which results from real-time electron oscillations in momentum space. In both situations, tr-RIXS exhibits particular advances in tracking the nonequlibrium behavior of multi-particle excitations, and is complementary to tr-ARPES and optics. With progress in experimental techniques as well as computing power and algorithms, we may address a wider range of systems including strongly correlated materials, topological materials, magnetic materials, and many others, and we believe tr-RIXS will bring new insights to the rich nonequilibrium physics in those systems.


\setcounter{secnumdepth}{0}
\section{Acknowledgements}
The work at Stanford University and SLAC National Accelerator Laboratory was supported by the US Department of Energy, Office of Basic Energy Sciences, Division of Materials Sciences and Engineering, under Contract No. DE-AC02-76SF00515. The work at Georgetown University was supported by the U. S. Department of Energy (DOE) Office of Basic Energy Science (BES) under Award number DE-FG02-08ER46542. In addition, Y.W. acknowledges support from a Postdoctoral Fellowship in Quantum Science from the Harvard-MPQ Center for Quantum Optics; J.K.F. was supported at Georgetown by the McDevitt bequest. This research used resources of the National Energy Research Scientific Computing Center (NERSC), a U.S. Department of Energy Office of Science User Facility operated under Contract No. DE-AC02-05CH11231.

\appendix*

\setcounter{equation}{0}
\setcounter{figure}{0}
\renewcommand\thefigure{A\arabic{figure}}
\section{Appendices}
\subsection{A: RIXS and Lindhard Response}\label{appendix:a}
\begin{figure*}[!t]
\begin{center}
\includegraphics[width=8.1cm]{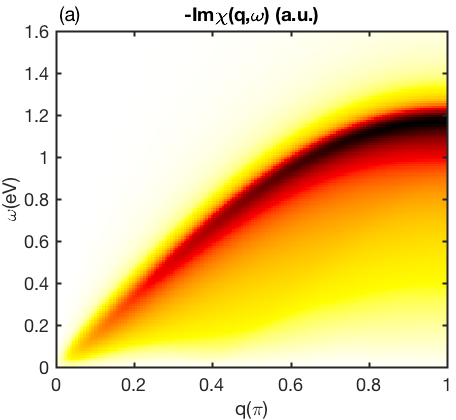}
\includegraphics[width=9.1cm]{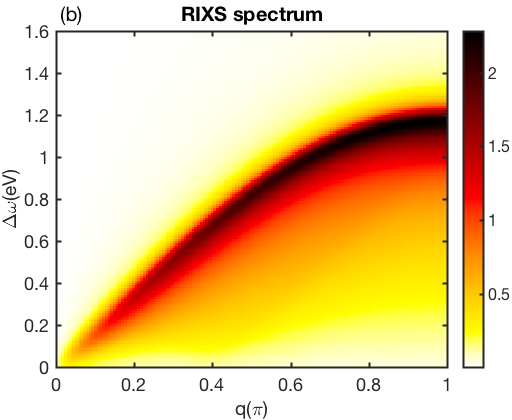}
\caption{\label{fig:A1}
Comparison of (a) the negative imaginary part of the Lindhard response calculated from Eq.\eqref{Lindhard} and (b) the equilibrium RIXS spectrum $I(\omega_{\rm i}, \Delta\omega, \bm{q})$ calculated from Eq.\eqref{eq} for the 2D tight-binding model Eq.\eqref{tight_binding} as parametrized in Sec.\ref{high_w}. In numerical evaluation, $\Gamma=0.80\textrm{eV}$ and $\delta=0.042\textrm{eV}$. $\bm{q}=(q,0)$ for both calculations.}
\end{center}
\end{figure*}

For the single-band model studied here, due to the full configuration of the core shell, the kinetic energy of the transient core hole vanishes. The Fourier transform of Eq.\eqref{A} yields (note that there are no spin flips)
\begin{equation}\label{A_k}
\begin{aligned}
A_{\rm f} &= \bra{\Psi^0_{\rm f}}\sum_{\bm{k}\sigma}p_{\bm{k}\sigma}\frac{1}{\omega_{\rm i}-\mathcal{H}+E_{\rm i}+i\Gamma}p_{\bm{k}+\bm{q} \sigma}^\dagger\ket{\Psi^0_{\rm i}}\\
&=\bra{\Psi^0_{\rm f}}\sum_{\bm{k}\sigma}p_{\bm{k}\sigma}G_c(\omega_{\rm i}-\epsilon_{\bm{k}+\bm{q}}, \bm{k}+\bm{q})p_{\bm{k}+\bm{q} \sigma}^\dagger\ket{\Psi^0_{\rm i}}\\
\end{aligned}
\end{equation}
where $p_{\bm{k},\sigma}$ annihilates a conduction electron with momentum $\bm{k}$ and spin $\sigma$, and $\epsilon_{\bm{k}}$ is its energy measured from the valence level in the atomic limit (i.e., $\epsilon_{\bm{k}}=\varepsilon_{\bm{k}}-\varepsilon_c-\mu_0$). Here, $G_c(\omega,\bm{p})=\frac{1}{\omega-\mu_0+i\Gamma}$ represents the core-hole propagator.
Thus
\begin{widetext}
\begin{equation}\label{eq}
\begin{aligned}
I(\omega_{\rm i},\Delta\omega,\bm{q})&\propto\sum_{\bm{k}}f(\epsilon_{\bm{k}})(1-f(\epsilon_{\bm{k}+\bm{q}}))G_c^*(\omega_{\rm i}-\epsilon_{\bm{k}+\bm{q}}, \bm{k}+\bm{q}) \delta(\Delta\omega-\epsilon_{\bm{k}+\bm{q}}+\epsilon_{\bm{k}})G_c(\omega_{\rm i}-\epsilon_{\bm{k}+\bm{q}}, \bm{k}+\bm{q})\\
&\propto-\Im\lim_{\delta\rightarrow 0^+}\sum_{\bm{k}}f(\epsilon_{\bm{k}})\big[1-f(\epsilon_{\bm{k}+\bm{q}})\big]\frac{1}{\Delta\omega-\epsilon_{\bm{k}+\bm{q}}+\epsilon_{\bm{k}}+i\delta}|G_c(\omega_{\rm i}-\epsilon_{\bm{k}+\bm{q}}, \bm{k}+\bm{q})|^2
\end{aligned}
\end{equation}

For the Lindhard response of such a system\citep{Mahan}, we have
\begin{equation}\label{Lindhard}
\begin{aligned}
\chi(\bm{q},\omega)=\frac{1}{V}\lim_{\delta\rightarrow 0^+}\sum_{\bm{k}}f(\epsilon_{\bm{k}})\big[1-f(\epsilon_{\bm{k}+\bm{q}})\big]\left[\frac{1}{\omega-\epsilon_{\bm{k}+\bm{q}}+\epsilon_{\bm{k}}+i\delta}-\frac{1}{\omega+\epsilon_{\bm{k}+\bm{q}}-\epsilon_{\bm{k}}+i\delta}\right]
\end{aligned}
\end{equation}
\end{widetext}
Note that the second term of $\chi(\bm{q},\omega)$ is negligible for positive $\omega$: either $f(\epsilon_{\bm{k}})\big[1-f(\epsilon_{\bm{k}+\bm{q}})\big]$ or $\Im\frac{1}{\omega+\epsilon_{\bm{k}+\bm{q}}-\epsilon_{\bm{k}}+i\delta}$ is close to zero. We can see that the RIXS spectrum is a convolution of the negative imaginary part of $\chi$ and $|G_c|^2$. We may compare the two numerically to see that they are basically the same, which is shown in Fig.\ref{fig:A1}.

\subsection{B: Floquet band renormalization}\label{appendix:b}
Under the steady-state assumption, the electronic wavefunction is determined by the Hamiltonian $\mathcal{H}_F$ with a periodic pump. We want to find the solutions to the Schr\"odinger equation
\begin{equation}\label{schrodinger}
i\frac{\partial}{\partial t}\ket{\psi(t)}=\mathcal{H}_F[A(t)]\ket{\psi(t)}
\end{equation} 
where $A(t+T)=A(t)$ and thus $\mathcal{H}_F(t+T)=\mathcal{H}_F(t)$. The Floquet theorem dictates that the generic solutions to this equation satisfy\citep{Floquet1883,Floquet1965}
\begin{equation}
\ket{\psi_\lambda (t)}=e^{-i\epsilon_\lambda T}\ket{u_\lambda (t)}
\end{equation}
where $\ket{u_\lambda (t)}=\ket{u_\lambda (t+T)}$.

We may expand both $\mathcal{H}_F(t)$ and $\ket{\psi(t)}$ in Fourier series as $\mathcal{H}_F(t) = \smashoperator{\sum\limits_{n=-\infty}^{\infty}} \mathcal{H}_n e^{in\Omega t}$ and $\ket{\psi_\lambda (t)} = \smashoperator{\sum\limits_{n=-\infty}^{\infty}} e^{-i\epsilon_\lambda t}e^{in\Omega t} \ket{u_{\lambda,n}}$. Then they are inserted into Eq.\eqref{schrodinger}, and we find solving $\epsilon_\lambda$ is equivalent to solving the following eigenvalue problem:
\begin{equation}\label{shorthand}
\sum_r \left( \mathcal{H}_{n-r}+n\Omega \delta_{n,r}\right)\ket{u_{\lambda,r}}=\epsilon_\lambda \ket{u_{\lambda,n}}
\end{equation}
\begin{figure*}[!ht]
\begin{center}
\includegraphics[width=12cm]{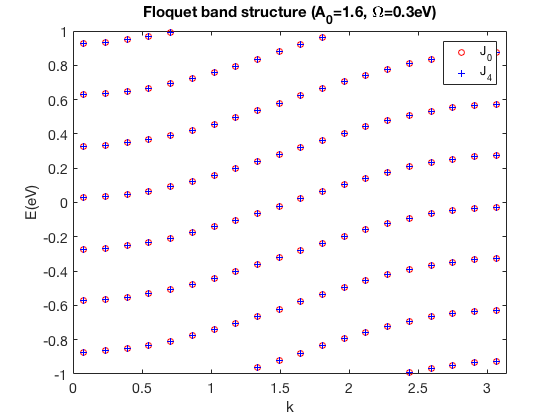}
\caption{\label{fig:A2}
The middle part (i.e., $n$ close to 0) of the Floquet band structure calculated from Eq.\eqref{shorthand} by only considering $\mathcal{J}_0$ and considering up to $\mathcal{J}_4$. In evaluating Eq.\eqref{shorthand} $r$ ranges from -12 to 12. Since the Fourier transform is linear, here we only considered terms in Eq.\eqref{tight_binding_no_mu} related to the $x$ direction (i.e., essentially a 1D tight-binding model). The two calculations are almost indistinguishable.}
\end{center}
\end{figure*}

Let $H_r$ be the matrix of $\mathcal{H}_r$. In explicit matrix form, we are solving the eigenvalues of the following infinite dimensional matrix:
\begin{equation}
H_F=
\begin{pmatrix}
\ddots & \ddots & \vdots & \vdots &  \\
\ddots & H_0+\Omega & H_1 & H_2 & \dots \\
\dots & H_{-1} & H_0 & H_1 & \vdots \\
\dots & H_{-2} & H_{-1} & H_0-\Omega & \vdots \\
 & \vdots & \vdots & \ddots & \ddots \\
\end{pmatrix}
\end{equation}

For the single-band tight-binding model Eq.\eqref{tight_binding} under a pump $\bm{A}(t)$, the Hamiltonian can be written as
\begin{equation}\label{tight_binding_no_mu}
\mathcal{H}(t)=\sum_{\bm{k}}\epsilon_{\bm{k}-\bm{A}(t)}p_{\bm{k}}^\dagger p_{\bm{k}},~ \epsilon_{\bm{k}}=-2t_h\left[\cos(k_x)+\cos(k_y)\right]
\end{equation}
and we have the following expansion\citep{shuxuewulifangfa}
\begin{align}
\cos&[k-A\cos(\Omega t)] = \mathcal{J}_0(A)\cos(k)\nonumber\\
&+\smashoperator{\sum_{m=1}^{\infty}}(-1)^m \mathcal{J}_{2m}(A)\cos(k)\left(e^{i2m\Omega t}+e^{-i2m\Omega t}\right) \nonumber\\
&+\smashoperator{\sum_{m=0}^{\infty}}(-1)^m \mathcal{J}_{2m+1}(A)\sin(k)\nonumber\\
&\ \ \ \ \ \ \ \ \times\big[e^{i(2m+1)\Omega t}+e^{-i(2m+1)\Omega t}\big]
\end{align}

Thus under a linear-polarized harmonic pump $\bm{A}(t)=\bm{A}_0\cos(\Omega t)$, by comparing coefficients of Fourier series, we may conclude that $H_r = \sum_{\bm{k}}E_{r,\bm{k}}p_{\bm{k}}^\dagger p_{\bm{k}}$, where

\begin{gather}
E_{0,\bm{k}} = -2t_h\left[\mathcal{J}_0(A_{x0})\cos(k_x)+\mathcal{J}_0(A_{y0})\cos(k_y)\right]\nonumber\\ 
E_{1,\bm{k}}=E_{-1,\bm{k}}=-2t_h\left[\mathcal{J}_1(A_{x0})\sin(k_x)+\mathcal{J}_1(A_{y0})\sin(k_y)\right]\nonumber\\
E_{2,\bm{k}}=E_{-2,\bm{k}}=2t_h\left[\mathcal{J}_2(A_{x0})\cos(k_x)+\mathcal{J}_2(A_{y0})\cos(k_y)\right]\nonumber\\
\cdots
\end{gather}

$H_F$ is a quasi-band matrix: when $|n|$ is large, $\mathcal{J}_n(A)\rightarrow 0$, $H_n\rightarrow \bm{0}$. As a result, the off-diagonal higher orders of $\mathcal{J}_n(A)$ contribute little to the quasienergy. Thus we only consider diagonal elements of $H_F$ in Sec.\ref{high_w} and obtain the quasienergy $\epsilon_{\lambda}=E_{\bm{k},n}=-2t_h\left[\mathcal{J}_0(A_{x0})\cos(k_x)+\mathcal{J}_0(A_{y0})\cos(k_y)\right]\pm n\Omega$. In fact, we have compared the quasienergy obtained considering up to $\mathcal{J}_4$ with that only considering $\mathcal{J}_0$, as shown in Fig.\ref{fig:A2}, with $A_0=1.6$ and $\Omega=0.3$eV. Note that $\mathcal{J}_4(1.6)=0.015$, which is already very small compared to $\Omega$ and $\mathcal{J}_0(1.6)=0.455$. The two calculations are almost indistinguishable, justifying the simplification in the band renormalization calculation. 
\bibliography{trRIXS}

\end{document}